# Design, fabrication, and testing of diamond axicons for X-ray microscopy applications

**Nazanin Samadi[1*], Frank Seiboth[2], Carlos Sato Baraldi Dias[3], Dmitri Novikov[1], Kathryn Spiers[1], and Xianbo Shi[4]**

[1] Deutsches Elektronen-Synchrotron DESY, Notkestr. 85, 22607 Hamburg, Germany
[2] Center for X-ray and Nano Science CXNS, Deutsches Elektronen-Synchrotron DESY, Notkestr. 85, 22607 Hamburg, Germany
[3] Institute for Photon Science and Synchrotron Radiation (IPS), Karlsruhe Institute of Technology (KIT), Hermann-von-Helmholtz-Platz 1, Eggenstein-Leopoldshafen, 76344, Germany
[4] Advanced Photon Source, Argonne National Laboratory, Lemont, IL 60564 USA

[*]E-mail: nazanin.samadi@desy.de

**Abstract.** This work presents the design, fabrication, and experimental validation of a refractive diamond axicon for X-ray beam shaping. The diamond axicon was developed to overcome the limitations of polymer-based axicons particularly for application in Transmission X-ray Microscopy (TXM) systems, offering superior mechanical strength, thermal stability, and radiation resistance, making it ideal for synchrotron applications. The axicon was fabricated using femtosecond laser ablation and tested at 11 keV under various coherence conditions. Results demonstrated that the axicon efficiently transformed the X-ray beam into a ring-shaped profile with over 80% transmission. Simulations confirmed the experimental findings and highlighted the potential for further improvements. This work paves the way for the use of diamond axicons in next-generation synchrotron facilities, with future efforts focusing on optimizing fabrication and testing the axicon in full TXM systems.

## 1. Introduction

Transmission X-ray Microscopy (TXM) [1-3] plays a critical role in high-resolution imaging across various fields, such as materials science, biology, and nanotechnology. The resolution of TXM is closely tied to the numerical aperture (NA) of both the condenser and the objective lens. The condenser has typically an annular aperture with a central stop in the middle. This annular aperture is covered by the divergence of the X-ray beam from third-generation sources. However, as synchrotron facilities upgrade to fourth-generation storage rings [4], the reduced X-ray beam divergence becomes less compatible with the large NA of the condenser typically used in TXM [5], creating new challenges.

To address this, new optical components that shape and condition the X-ray beam before it reaches the condenser are required. An axicon, a conical optical element, can transform the beam into a uniform ring-shaped profile, ideal for matching the condenser's acceptance typically used in TXM setups. Our previous work [5] demonstrated the feasibility of a 3D-printed polymer axicon for this purpose, but its susceptibility to radiation damage highlighted the need for more robust materials.

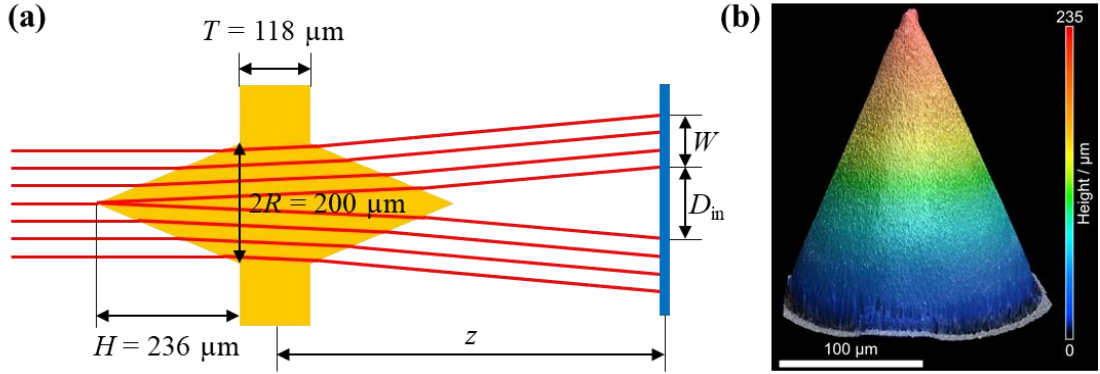

**Figure 1.** (a) Schematic view of the double-sided diamond axicon design. (b) A confocal laser scanning microscope image, showing the surface of one of the axicon cone.

Diamond's exceptional mechanical strength, thermal stability, and resistance to radiation make it an ideal candidate for X-ray optics in modern synchrotrons. In this paper, we present the development, manufacturing, and testing of a refractive diamond axicon designed to address the limitations of the radiation hardness of the previous axicon [5]. The diamond axicon was designed following the approach in [5] and manufactured to micron accuracy using laser ablation [6], resulting in precise structures optimized for beam shaping. Experimental results demonstrate the axicon's ability to produce high-quality ring-shaped beam profiles, which were validated by simulations. This advancement in optical components for TXM systems paves the way for enhanced imaging capabilities in next-generation synchrotron facilities.

## 2. Design and Fabrication of Diamond Axicons

Following the procedures outlined in Ref. [5], we designed a diamond axicon with a double-sided cone geometry, as shown in Fig. 1(a). The design parameters were chosen to create a ring-shaped beam profile at 11 keV, with an inner diameter $D_{in}$ = 200 μm at a distance $z$ = 3.5 m. The axicon cone radius was set to $R$ = 100 μm. For a diverging beam with a source-to-axicon distance of $p$ = 110 m, the ring width at the distance z is given by $W = R(1+z/p)$ = 103 μm. The cone height $H$ is calculated to be 236 μm using the following equation [5]:

$$H = \frac{D_{in} R}{2 N z \delta}, \quad (1)$$

where $N$ represents the number of axicons, here $N$=2 because we used a double-sided axicon, and $\delta$ is the refractive index decrement. The remaining diamond substrate thickness $T$ = 118 μm can be further reduced in future designs to minimize absorption.

The diamond axicon was fabricated using femtosecond laser ablation [6]. The ultrafast laser is a diode-pumped fiber laser (Amplitude Satsuma HP2) with 1030 nm and 285 fs pulse length. For ablation, we used the third harmonic at 343 nm with an NA 0.36 objective. A confocal laser microscope (Keyence VK-X110) image of the surface of one of the axicon cones is presented in Fig. 1(b).

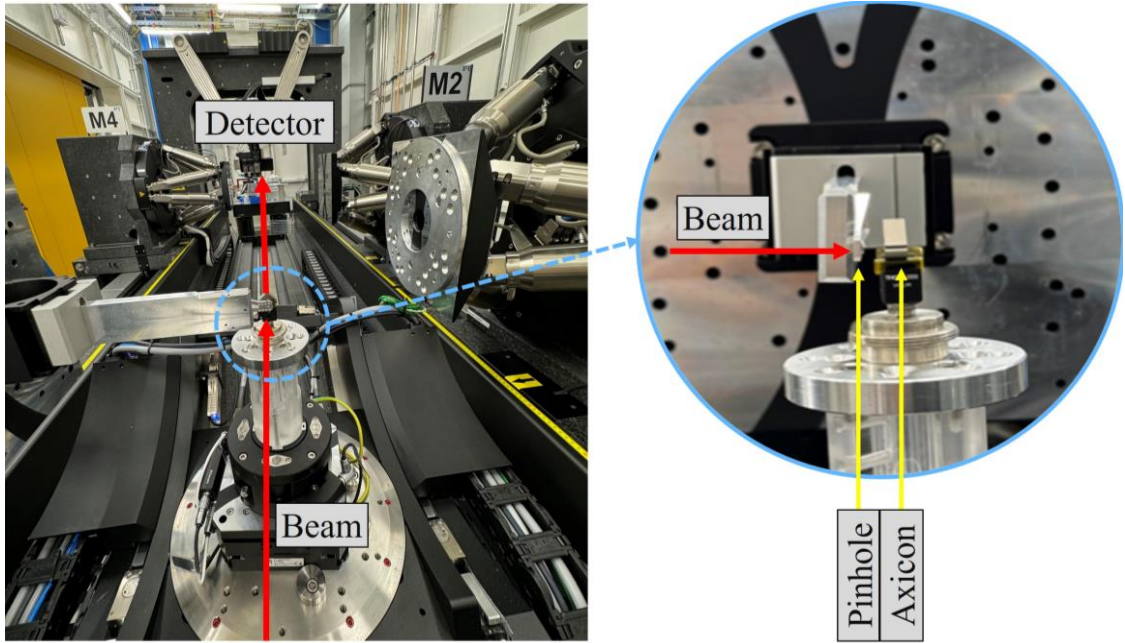

**Figure 2.** Photo of the diamond axicon experimental setup at the HIKa beamline.

### 3. X-ray experimental setup

The experimental test was carried out at the HIKA instrument on the P23 beamline at PETRA III. The beam was generated by a U32 undulator with a 31.4 mm period and 63 periods, situated in the high-beta section of the storage ring. The Si (111) double crystal monochromator was tuned to 11 keV. The total source dimensions, accounting for both the electron beam and undulator radiation contributions, are as follows: the sigma source sizes are $\Sigma_x$ = 141.5 µm and $\Sigma_y$ = 5.95 µm, while the sigma divergences are $\Sigma'_x$ = 8.86 µrad and $\Sigma'_y$ = 5.71 µrad. A slit with a horizontal aperture positioned at about $z_s$ = 71 m from the source was used to control the beam's degree of coherence.

The diamond axicon was placed 110 meters from the source, with a 200 µm diameter pinhole upstream to match its acceptance aperture. The detector, located at $z$ = 3.4 meters (slightly offset from the design value) downstream of the axicon, consisted of an sCMOS PCO edge camera paired with a 10× objective lens, providing an effective pixel size of 0.65 µm. Each image presented in this work represents the average of 50 acquisitions unless otherwise noted. The acquisition time was adjusted based on varying conditions to ensure a sufficient signal-to-noise ratio.

### 4. Results

Beam images with the pinhole only, and with both the pinhole and the diamond axicon, were measured under different coherence conditions, as summarized in Fig. 3. To achieve a highly coherent illumination on the axicon, as shown in Figs. 3(a) and (b) a slit with a horizontal aperture of 50 µm 71 meters upstream from the source was used. The transverse coherence length in the vertical direction was sufficiently large and required no additional aperture. The partially coherent case was obtained with the horizontal slit aperture fully open. To generate the incoherent case, a decoherer was introduced upstream of the axicon setup, consisting of a piece of sandpaper attached to a continuously rotating motor (a drill).

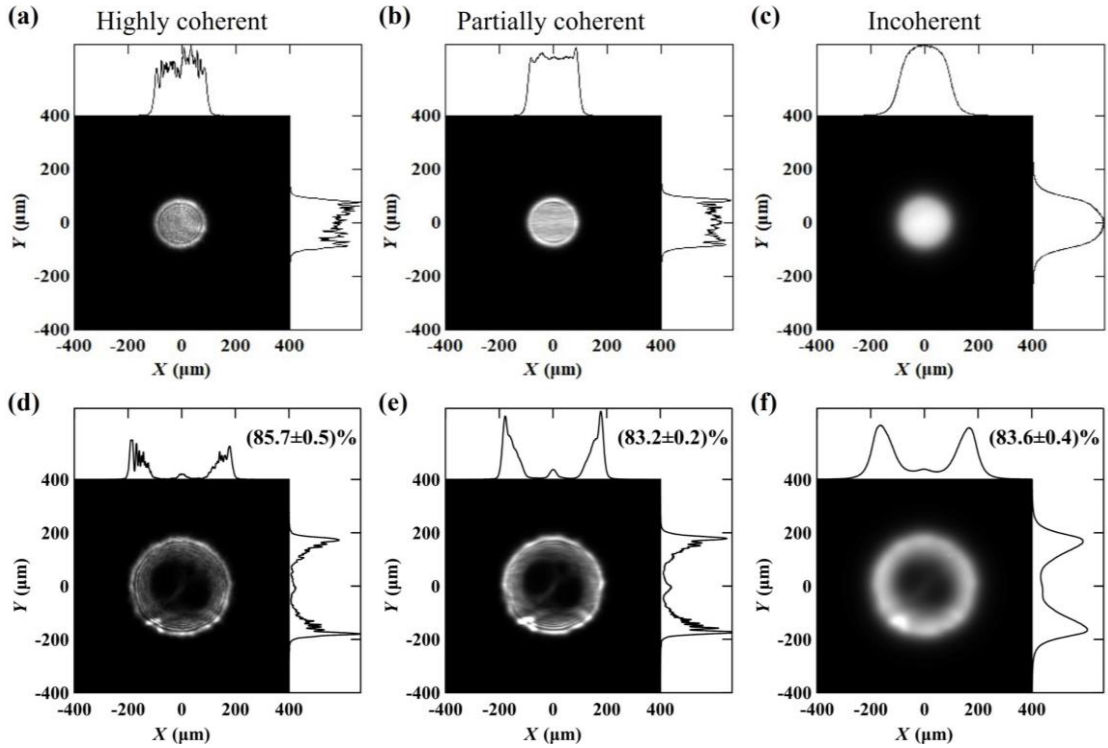

**Figure 3.** Beam intensity on the detector along with horizontal and vertical line profiles with the pinhole only (a-c) and with both the pinhole and the diamond axicon (d-f) at 11 keV under different coherence conditions; see text for more details. The transmission values extracted from the measurements are indicated in figures (d-f).

The use of the diamond axicon successfully reshape the circular beam through the pinhole to a ring-shaped profile in all cases. Under the highly coherent condition, diffraction fringes from the pinhole aperture are clearly visible along both transverse directions, as shown in Figs. 3(a) and 3(d). When the horizontal slit was opened to reduce the horizontal coherence, the fringe contrast along the horizontal direction decreases, as observed by comparing Figs. 3(b) and 3(e) with Figs. 3(a) and 3(d), respectively. Once the decoherer was employed, the images in Figs. 3(c) and 3(f) show the smeared profiles, where the more uniform beam could provide better matching with the TXM condenser.

The transmission values of the diamond axicon were calculated as the ratio of the integrated image intensities between the configurations with both the pinhole and axicon, and those with the pinhole only. An overall transmission of greater than 80% shows promise for a significant gain inl TXM applications. However, some non-uniformity in the beam profiles was observed, likely due to a combination of axicon fabrication imperfections and non-uniformities in the incoming beam, which is influenced by all upstream optics. Further improvements in diamond axicon fabrication are currently underway.

The intensity profiles were simulated under the same conditions as those in the experiments, with the results shown in Fig. 4. The simulation was performed using near-field, i.e. Fresnel, wavefront propagation, as detailed in ref [5], assuming an aberration-free axicon. The simulated beam profiles and coherence effects closely match those observed in the experimental results, and the simulated transmission values are in good agreement with the measured values. This validation further confirmed the success of the diamond axicon as a promising radiation-resistant option for broader TXM applications. It is worth noting that the transmission of the axicon can be improved significantly by reducing the diamond substrate thickness $T$, as indicated in Fig. 1(a). At 11 keV, reducing $T$ from 118 µm to 50 µm could increase the transmission by approximately 4%.

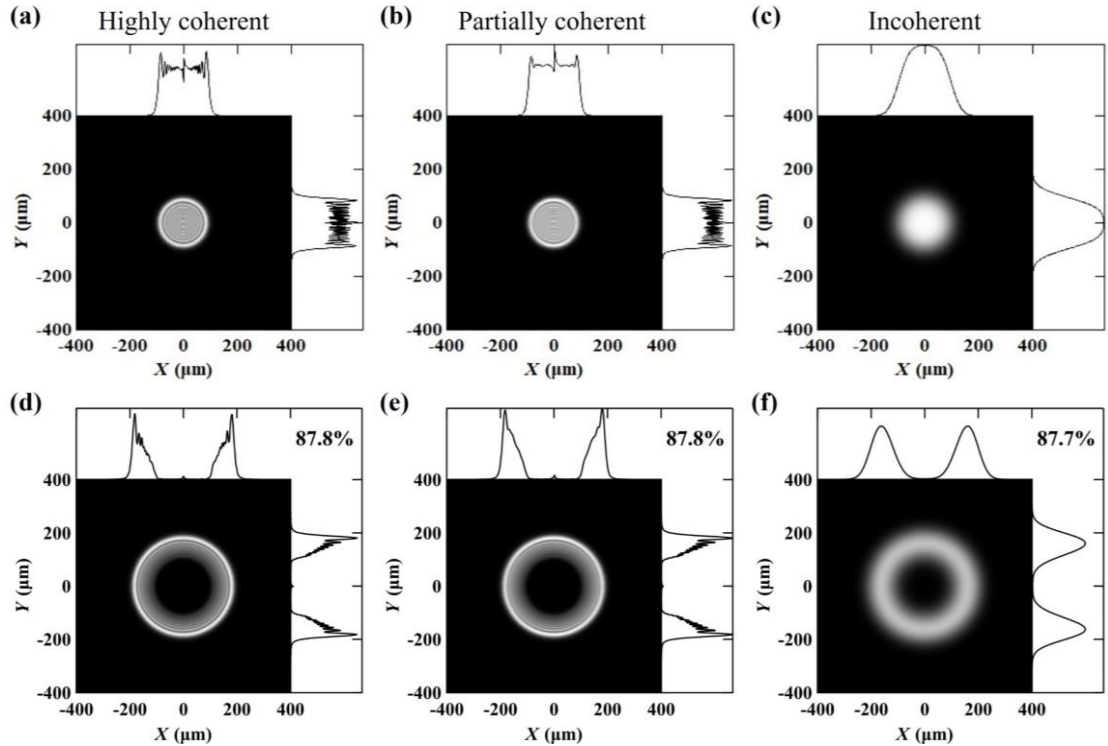

**Figure 4.** Simulated beam images under the same conditions as those in Fig. 3.

## 5. Conclusion

This work demonstrated the successful development of a refractive diamond axicon for X-ray beam shaping, addressing the limitations of polymer-based axicons in high-radiation environments. The diamond axicon achieved over 80% transmission and generated high-quality ring-shaped beam profiles under various coherence conditions, proving its suitability for synchrotron applications. Simulation results further validated the experimental findings, confirming the axicon's potential for broader use in TXM and similar optical systems.

Future work will focus on improving the fabrication quality to reduce imperfections, which will help produce more uniform beams. Additionally, we aim to enhance the transmission efficiency by reducing the diamond substrate thickness. Testing the axicon in a full TXM system will also be a key step, alongside comparing different types of axicons to further optimize TXM performance for next-generation synchrotron applications.

**Acknowledgments**

We acknowledge DESY (Hamburg, Germany), a member of the Helmholtz Association HGF, for the provision of experimental facilities. Parts of this research were carried out at the HIKa instrument of beamline P23, PETRA III.